# GPT-4o reads the mind in the eyes

James W. A. Strachan[1], Oriana Pansardi[1,2,3,#], Eugenio Scaliti[1,4,5,#], Marco Celotto[6,7], Krati Saxena[8], Chunzhi Yi[6,9], Fabio Manzi[8], Alessandro Rufo[8], Guido Manzi[8], Michael S. A. Graziano[10], Stefano Panzeri[6,*], Cristina Becchio[1,*]

1. Department of Neurology, University Medical Center Hamburg-Eppendorf (UKE), Hamburg, Germany
2. Cognition, Motion and Neuroscience, Italian Institute of Technology, Genoa, Italy
3. Department of Psychology, University of Turin, Turin, Italy
4. Department of Management "Valter Cantino", University of Turin, Turin, Italy
5. Human Science and Technologies, University of Turin, Turin, Italy
6. Institute for Neural Information Processing, Center for Molecular Neurobiology (ZMNH), University Medical Center Hamburg- Eppendorf (UKE), Hamburg, Germany
7. Department of Brain and Cognitive Sciences, Picower Institute for Learning and Memory, Massachusetts Institute of Technology, Cambridge, MA, USA
8. Indeep Artificial Intelligence, Las Palmas de Gran Canaria, Spain
9. The School of Medicine and Health, Harbin Institute of Technology, Harbin, Heilongjiang, China
10. Princeton Neuroscience Institute, Princeton University, Princeton, NJ, USA

\# These authors contributed equally
\* Correspondence: c.becchio@uke.de, s.panzeri@uke.de

**Abstract**

Large Language Models (LLMs) are capable of reproducing human-like inferences, including inferences about emotions and mental states, from text. Whether this capability extends beyond text to other modalities remains unclear. Humans possess a sophisticated ability to read the mind in the eyes of other people. Here we tested whether this ability is also present in GPT-4o, a multimodal LLM. Using two versions of a widely used theory of mind test, the Reading the Mind in Eyes Test and the Multiracial Reading the Mind in the Eyes Test, we found that GPT-4o outperformed humans in interpreting mental states from upright faces but underperformed humans when faces were inverted. While humans in our sample showed no difference between White and Non-white faces, GPT-4o's accuracy was higher for White than for Non-white faces. GPT-4o's errors were not random but revealed a highly consistent, yet incorrect, processing of mental-state information across trials, with an orientation-dependent error structure that qualitatively differed from that of humans for inverted faces but not for upright faces. These findings highlight how advanced mental state inference abilities and human-like face processing signatures, such as inversion effects, coexist in GPT-4o alongside substantial differences in information processing compared to humans.

**Introduction**

The diffusion of large language models (LLMs) has enabled a novel type of language game[1]: prompting LLMs with amusing probes to reveal their hollowness. In one such game, when Douglas Hofstadter asked GPT-3, "What's the world record for walking across the English Channel?", GPT-3'replied, "The world record for walking across the English Channel is 18 hours and 33 minutes"[2]. This exchange has been cited as evidence that LLMs are fundamentally clueless. However, an alternative perspective suggests that GPT-3 might have recognized the absurdity of the question and responded playfully, mirroring how a human would respond[3]. This raises the question: can LLMs perceive human mental states? Could GPT-3 have sensed the amusement of its interrogator?

While GPT-3 was limited to text-based processing and could not interpret human expressions, the development of multimodal models like GPT-4o moves us closer to a future where LLMs might not only generate responses but also recognize and react to human mental states. This could entirely change the nature of human-AI interaction, enabling LLMs to perceive and rapidly react to tacit social information, anticipate human behaviour, and adapt dynamically to the social environment. To achieve this capability, LLMs would need to reproduce human's ability to read mental states from subtle variations in behaviour — a raised eyebrow, a smirk, or a frown.

Substantial prior research has shown that LLMs can attain human-like performance in cognitive tasks using text inputs. They can reproduce human-like decisions and deliberations[4], generate human-like inferences about emotions[5], and even generate human-like inferences about sophisticated mental states[6]. By contrast, their abilities to process nontextual stimuli remain poorly explored. Previous research[7] found that perceptual evaluation of social features in dynamic stimuli is similar in GPT-4V and humans, suggesting that GPT-4V has human-like social perceptual capabilities. However, the extent to which these capabilities support mentalistic inferences is still unclear.

Here, we applied one of the most widely used tests in the Theory of Mind literature[8], the Reading the Mind in the Eyes Test (RMET) (Baron-Cohen et al., 2001), to compare the mindreading ability of GPT-4o against human evaluations. Humans can read highly complex mental states from the eyes of other humans. The RMET is an advanced theory of mind task designed to measure this ability [9]. In this paper-and-pencil (now online) test, responders are presented with pictures of the eye region of a human face and asked to indicate which of the four-word choices best describes what the person in the picture is thinking or feeling (Figure 1A). The test was designed to have sufficient analytical complexity to allow assessment of individual differences in social cognition in a human adult population[9] and is one of two tests recommended for "understanding mental states" by NIMH Research Domain Criteria (https://www.nimh.nih.gov/research/research-funded-by-nimh/rdoc/constructs/understanding-mental-states accessed on October 20, 2024). Since its publication, it has been employed to assess theory of mind in adults with and without psychopathology, brain damage, and dementia[9-13]. Here, we used the RMET and the Multiracial RMET (MRMET)[14], a validated alternative to the RMET consisting of racially inclusive, non-gender-stereotyped stimulus faces, to compare human and GPT-4o's eye reading abilities (Figure 1A).

A classic signature of human face processing is the inversion effect, where perceptual judgments about face identity[15] and emotion[16,17] are more difficult when faces are inverted compared to when they are upright. This effect has been instrumental in investigating distinct modes of processing — configural versus featural[18-20], holistic versus part-based[21] — while controlling for low-level properties of face images. Here, we leveraged this effect to reveal commonalities and differences in the processing of mentalistic information between humans and GPT-4o. Applying concepts from information theory to characterize the error spaces of humans and GPT-4o enabled us to show that, while inversion induces quantitative changes in information processing humans, it induces qualitative changes in GPT-4o.

**Results**

**Reading the Mind in the Eyes Test (RMET)**

The RMET consists of 36 photographs displaying the eye region of faces. For each image, responders are asked to choose the word from a set of four options that best describes what the person in the photograph is thinking or feeling (Figure 1A). To compare the use of information in humans and GPT-4o, we manipulated the orientation of images to be either upright or inverted and administered the test to GPT-4o (25 repetitions for each orientation) and a sample of 51 human adults. Accuracy, measured as fraction correct, is shown in Figure 1B. Mixed-effect logistic regression with model (human, GPT-4o) and orientation (upright, inverted) as fixed effects and observation (subject ID for humans, session ID for GPT-4o) as a random intercept, revealed a significant main effect of model and significant main effect of orientation (Table S2). These effects were further qualified by a significant model-by-orientation interaction, reflecting the fact that GPT-4o performed significantly better than humans when reading upright stimuli but significantly worse than humans when reading inverted stimuli (Table S3). Numerically, humans experienced a decrement of about 15% (odd ratio, OR = 1.99) for inverted stimuli, which aligns with the findings from previous studies[22]. In contrast, GPT-4o showed a decrement exceeding 50% (OR = 11.67). While it is acknowledged that GPT models have limitations with rotated images (https://platform.openai.com/docs/guides/vision accessed on August 6, 2024), this pattern of results could be interpreted as evidence that inversion completely disrupted GPT-4o ability to extract diagnostic information from faces. However, one reason for caution in interpreting this pattern of results is that with over 8,000 citations to date since its publication and an annual citation rate of over 500 (Google Scholar), the RMET is one of the most widely used tests in the Theory of Mind literature. This raises the concern that the faces (which were cut from magazines), and the response options may have been included in the training dataset for GPT-4o, potentially inflating its performance for upright faces. Another concern about the RMET is that all the 36 stimuli are of white faces. This raises the question of whether these results would generalize to non-white faces.

**Multiracial RMET (MRMET)**

To address concerns regarding exposure to test stimuli during training and examine the generality of the effects obtained in the RMET, in a second set of experiments we tested the ability of GPT-4o to identify mental states from a new set of items, including both white and non-white face stimuli (MRMET)[14]. The MRMET was first published online in April 2024, ruling out that it could have been included in the materials used to train GPT-4o, which was released in May 2024 and trained on material dating to October 2023 (https://platform.openai.com/docs/models/gpt-4o accessed on August 6, 2024). Using the same design as for the RMET, we administered the MRMET to GPT-4o and humans, increasing the number of repetitions to 200. Accuracy, measured as fraction correct, is shown in Figure 1C. Analysis using mixed-effect logistic regression replicated the pattern observed in the RMET. We observed significant main effects of model and orientation, both qualified by a significant model-by-orientation interaction (Table S2). As shown in Figure 1C, GPT-4o outperformed humans in recognizing mental states from upright faces but showed a significant lower performance than humans when faces were inverted (TableS3). While humans showed a significant decrease in accuracy for inverted faces, they still performed above chance. In contrast, GPT-4o's accuracy for inverted faces dropped below chance level.

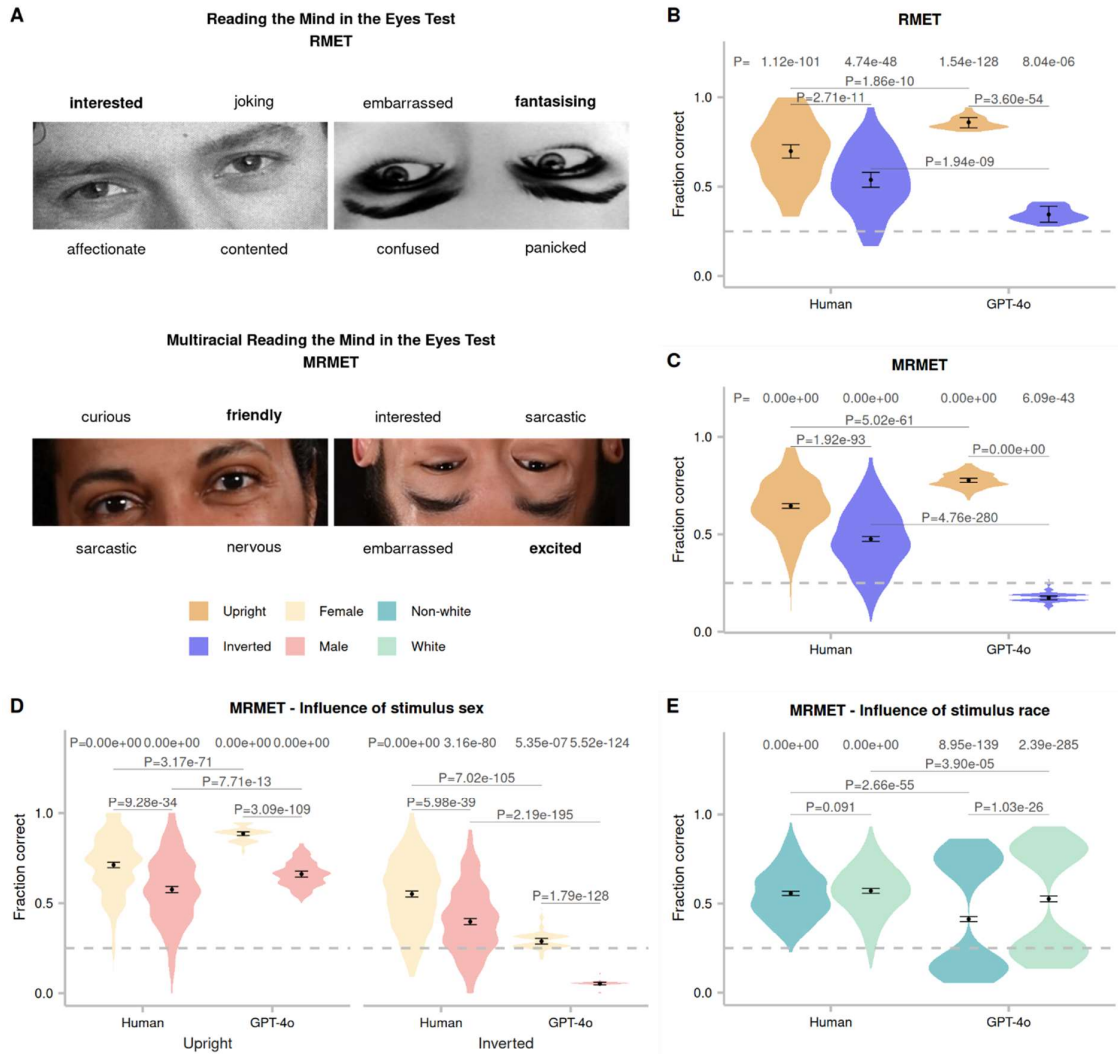

**Figure 1**. **Accuracy of humans and GPT-4o the Reading the Mind in the Eyes Test (RMET) and the Multiracial Reading the Mind in the Eyes Test (MRMET). A.** Example upright and inverted stimuli and response alternatives for RMET and MRMET. The correct response alternative is printed in bold font. **B.** Distributions of fraction correct on individual runs of the test for humans (N = 51) and GPT-4o (n = 25 repetitions for each orientation) on the upright and inverted orientations of the RMET. **C.** Distributions of fraction correct on individual runs of the test for humans (N = 400) and GPT-4o (n = 200 for each orientation) on the upright and inverted orientations of the MRMET. **D.** Distributions of fraction correct on individual runs of the MRMET for humans and GPT-4o for female and male stimuli broken down by orientation. **E.** Distributions of fraction correct on individual runs of the MRMET for humans and GPT-4o for White and Non-White stimuli. In panels B-E, violin plots show the distribution of fraction correct, black dots indicate the estimated marginal means (EMMs) estimated from logistic regression models, and error bars show the 95% confidence intervals of the EMMs. *P* values with brackets correspond to the planned contrasts of the EMMs, while *P* values along the top of each plot correspond to tests of each EMM against chance (0.25). All *P* values are two-tailed, and Holm-Bonferroni corrected.

**Effect of stimulus race.** The MRMET includes White and Non-white stimuli. To examine the effect of race on mental state attribution, we repeated our analysis including race (White, Non-white) as a fixed effect. This analysis revealed a significant model by stimulus race interaction (Table S2), reflecting GPT-4o's superior performance with White stimuli compared to Non-white stimuli, whereas humans showed no significant difference between White and Non-white stimuli (see Figure 1E and Table S3). In humans, stimulus race has been reported to interact with the race of the participants, with Non-white participants showing an advantage for Non-white stimuli[14]. To assess whether a similar same-race advantage was present in our human data, we conducted a mixed-effect logistic regression with participant race (White, Non-White), stimulus race (White, Non-White), and orientation (upright, inverted). This analysis revealed a three-way interaction between participant race, stimulus race, and orientation (Table S2), reflecting a same-race advantage in White participants for White inverted over Non-white inverted stimuli but no same-race advantage for upright stimuli.

**Effect of stimulus sex.** We investigated the effect of stimulus sex (female, male) by conducting a mixed-effects logistic regression with model, orientation, and sex as fixed effects. This analysis revealed a significant interaction among model, orientation, and sex (Table S2). Breaking down this three-way interaction by orientation showed that while both humans and GPT-4o performed better for female stimuli, inversion disproportionately affected GPT-4o's processing of male stimuli, causing performance to drop below chance level (Figure 1D, Table S3). Previous studies in humans reported an on-average advantage for female participants over male participants[8,14]. In our human data, we found no significant female advantage (Table S2). However, there was a significant interaction of orientation by subject sex (Table S2), reflecting a stronger modulation by sex for inverted stimuli than upright stimuli (Table S3).

Together with the analysis of the RMET, these findings demonstrate that GPT-4o outperformed humans when reading mental states from upright faces and exhibited a vulnerability to face inversion that was even more pronounced than in humans.

**The argument from error**

Accuracy, measured as fraction correct, is a useful metric, but it provides only a partial perspective on the computations involved in extracting information from face images. Because there are more ways to be wrong than to be right, analysing the pattern of errors across different conditions can reveal processing differences that are not detectable by comparing the fraction of correct answers. To investigate these differences, we next compared the error space of GPT-4o and humans. Figure 2A presents confusion matrices for human and GPT-4o responses to upright and inverted stimuli[23,24]. These matrices report the conditional probability of the reported mental state given the presented mental state, for all combinations of presented and reported mental states (Table S7). The on-diagonal elements represent correct classifications, while the off-diagonal elements show the patterns of misclassification (errors).

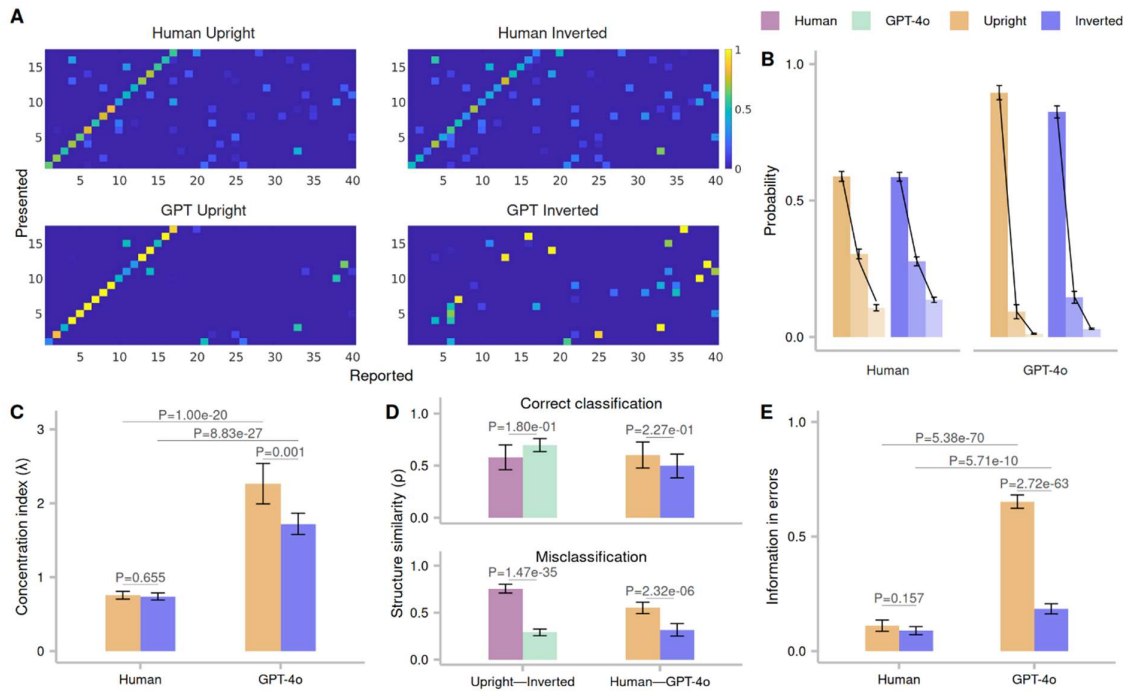

**Figure 2. Analysis of human and GPT-4o errors on the MRMET. A.** Confusion matrices for human and GPT-4o for upright and inverted stimuli. Presented mental state label (i.e., the true mental state posed by the face) is shown on the y-axis (1-17) and reported mental state label (the mental state identified by the participants among the four response options) on the x-axis (1-40). 18-40 on the x-axis correspond to labels that were never presented, and the corresponding matrix entries are thus not plotted (see Table S7). Cell colour indicates the conditional probability of the reported mental state given the presented mental state. **B.** Probabilities of the three incorrect alternatives ranked in terms of decreasing probability of being chosen. Black lines show exponential decay function **C.** Concentration index (λ) for each condition, which measures the rate of decay of probabilities across alternatives such that higher values indicate greater concentration on a single alternative, while lower values indicate more dispersion across alternatives. **D.** Structure similarity computed as Spearman correlation coefficient between the following vector pairs: human upright and inverted, GPT-4o upright and inverted, human upright and GPT-4o upright, and human inverted and GPT-4o inverted. Correlations are shown separately for correct classifications (top row) and misclassifications (bottom row). **E.** Information in errors. Bars show mean and 95% confidence intervals calculated by bootstrap. In panels C-E, pairwise comparisons report the Holm-Bonferroni corrected *P* values of independent two-tailed t-tests.

**Error concentration.** As a first step to quantify differences in error patterns, we computed, separately for each model and orientation, the distribution of errors across response options. As in the RMET, in the MRMET, each face is presented with a set of four response options, one correct option and three incorrect options. If response were random – irreproducible across individual runs and unrelated to the presented mental state – each incorrect options would have a 1/3 probability to be chosen on incorrect trials. The degree to which the observed distribution of errors deviates from this equiprobable distribution provides a measure of how consistent errors are across individual runs. Figure 2B and Table S4 show the probabilities of error, averaged across all possible represented mental states, for GPT-4o and humans, to upright and inverted faces. For both orientations, GPT-4o errors were far more consistent than human errors. To quantify this consistency, we fitted each error distribution to an exponential probability decay function, which assumes that the probability of selecting an incorrect option decreases at a constant rate between the first and second, and second and third alternatives. This exponential decay function is described by a single concentration index, $\lambda$, which measures the rate of decay. Lower values of this index indicate more dispersion among alternatives (less concentration), higher values indicate less dispersion (more concentration). Fitting $\lambda$ to response option probabilities revealed that the concentration of errors did not significantly differ across orientations for humans, but it did for GPT-4o, with responses to upright faces being more tightly concentrated than responses to inverted faces (Figure 2C, Table S5). Across both orientations, GPT-4o errors were significantly more concentrated than human errors (Figure 2C, Table S5). These findings indicate a higher consistency of GPT-4o's errors relative to human errors.

**Similarity in structure.** Error distribution analysis indicates a more consistent error pattern for GPT-4o relative to humans. However, this analysis does not reveal the degree of similarity between the response patterns of humans and GPT-4o. To quantify similarities between models and orientation, we next flattened each confusion matrix into a vector and computed the Spearman correlation coefficient between human and GPT-4o upright and inverted vectors separately for correct classifications (on-diagonal elements in the confusion matrices) and misclassifications (off-diagonal). Similarity was defined as the correlation coefficient between vectors.

As shown in Figure 2D, for on-diagonal elements, similarity was moderate and stable across models and orientations (Table S6). Specifically, similarity in the pattern of correct classifications did not differ between human upright and human inverted conditions, nor between GPT-4o upright and GPT-4o inverted conditions. It also did not differ between human upright and GPT-4o upright, or between human inverted and GPT-4o inverted (Table S6). These findings indicate that correct responses tended to be similar between orientations within each model and between human and GPT-4o models.

In contrast, for the off-diagonal elements (misclassifications), a divergent pattern emerged. While humans exhibited a strong similarity between errors for upright and inverted stimuli, GPT-4o showed only weak similarity between errors in the two orientations (Figure 2D). Additionally, the similarity between errors for human inverted and GPT-4o inverted was weak and significantly lower than the similarity between human upright and GPT-4o upright (Table S6). For example, humans confused 'threatening' (label 17) with 'angry' for both upright and inverted orientations, GPT-4o labelled it as 'angry' in incorrect upright trials but systematically confused it with 'scared', the least likely error for humans, in inverted trials. These findings suggest a substantial dissimilarity in GPT-4o's error pattern for inverted images, both when compared to its own pattern for upright images and to the pattern of humans for inverted images.

**Information in errors.** How errors are structured can provide information about mental states not captured by the fraction of correct responses. To carry information an error pattern must be consistent and able to differentiate between the presented mental state[25,26]. To quantify information carried by error patterns in our data, we first computed the mutual information between the rows and columns of the confusion matrix for each condition[23,24]. The mutual information calculated in this manner provides a measure of the total information carried by both correct and incorrect classifications. Next, we subtracted the information carried by a confusion matrix in which correct responses (on-diagonal elements) are preserved but off-diagonal elements are flattened for each mental state (minimum concentration). This adjustment disrupts information carried by the error pattern. By dividing the resulting difference by the maximal information that could be added by the error structure (see Methods), we obtained a normalized index of the information specifically carried by the pattern of errors (Information in errors), above and beyond the information carried by correct responses. A value of 0 of this index indicates that errors are equiprobably distributed and unrelated to the presented mental state, providing no additional information. A value of 1 indicates that errors are maximally informative, being concentrated on the incorrect option that, among the empirically possible alternatives, maximally differentiates between presented mental states.

The results of this analysis are shown in Figure 2E (Table S5). Across both orientations, GPT-4o's error pattern contained significantly more information than the human error pattern (Table S5). While the information content of human errors did not differ between orientations, the information content of GPT-4o's errors was lower for inverted stimuli compared to upright stimuli (Table S5). These findings complement the error similarity results, suggesting that both the structure and information content of human error pattern remained stable across orientations. This indicates that when humans made mistakes, they did so more or less randomly, regardless of stimulus orientation. GPT-4o, on the other hand, exhibited a more structured error pattern than humans, with an orientation-dependent structure that carried significantly higher information content for upright stimuli compared to inverted stimuli.

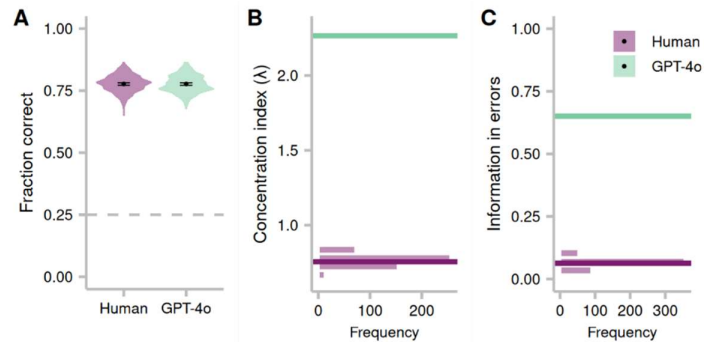

**Figure 3. Human responses on upright faces sampled from the MRMET validation dataset[14] to match the distribution of fraction correct of GPT-4o. A.** Example human sample (N = 200) drawn from the MRMET validation dataset[14] to match the mean and SD of GPT-4o fraction correct. Violin plots represent the distribution of fraction correct. Black dots indicate the mean. Error bars show the 95% confidence intervals for the fraction correct of the two samples. **B.** Distribution of the concentration index computed across 500 sampled subsets from the MRMET validation dataset, plotted against the concentration index of GPT-4o responses (green line). **C.** Distribution of information in errors computed across 500 sampled subsets from the MRMET validation dataset, plotted against information in errors of GPT-4o responses (green line). The dark purple lines in panel B-C represent the values of the concentration index and information in errors for our human data and reported in Figures 2C and 2E.

**Influence of the variability of performance on information**

Humans showed considerably higher inter-individual variability in performance compared to GPT-4o (Figure 1C). To control for the influence of this variability on the differences in the consistency and information content of errors, in a control analysis, we subsampled groups of human participants (500 samples of N = 200 human subjects) from the validation dataset of the MRMET[14] (N = 9295) to match the mean and standard deviation of fraction correct of the GPT-4o sample (see Figure 3A for an example sample). Even after matching performance, GPT-4o showed much higher consistency (Figure 3B) and information content of errors (Figure 3C) compared to humans. The distribution of the concentration index and information in errors computed across these performance-matched human samples was closely distributed around the values of these indices observed with our original human sample. This suggests that the differences in consistency and information content between GPT-4o and humans are not due to performance variability but reflect fundamental differences in how information is processed.

**Discussion**

Psychology researchers have developed reliable instruments for evaluating social cognition and social processing in both the laboratory and the clinic[9]. The present study sought to examine the ability of large language models to infer mental states from pictures of the eye region comparing GPT-4o's performance to that of humans using two validated mind-reading tests, the Reading the Mind in the Eyes Test (RMET) and the Multiracial RMET (MRMET). Our findings provide novel insights into GPT-4o's social sensitivity and highlight key differences in mindreading by GPT-4o and humans.

The RMET was designed to have sufficient analytical complexity to allow assessment of individual differences in social cognition in an adult human population. A high score on this test indicates the ability to judge mental states extracting information from minimal features around the eyes. GPT-4o's impressive performance on both the RMET and the MRMET demonstrates its capability to extract information about mental states from these minimal features.

A classic signature of human face processing is face inversion effect[15,27]. Both human accuracy and GPT-4o's accuracy showed this effect. However, GPT-4o's accuracy was significantly more affected by inversion compared to humans. This heightened vulnerability to inversion likely reflects GPT-4o's much higher exposure to upright faces relative to inverted faces during training. Experience-dependent deficits for inverted face have been reported in convolutional neural networks (CNNs) trained on upright faces[28]; these effects reverse when CNNs are trained on inverted stimuli[28]. In LLMs, the inversion effect may spontaneously arise as a result of reduced exposure to inverted faces during training.

Despite GPT-4o's heightened vulnerability to inversion, its error space remained highly structured even for inverted stimuli. Our error analysis reveals both commonalities and differences between the error spaces of GPT-4o and humans for upright and inverted stimuli. Humans displayed an error pattern that, while only moderately consistent across individual runs, was highly reproducible across upright and inverted orientations. In contrast, GPT-4o exhibited an error pattern that was highly consistent across runs within each orientation but showed substantial differences between orientations.

Applying concepts from information theory enabled us to show that information content of human and GPT-4o error patterns was also markedly different. The capacity of an error pattern to carry information depends both on its consistency and its capacity to differentiate between presented mental states. As such, information in errors can reveal whether mental-state-related information is extracted in incorrect trials and how consistently this information is utilized. The high information content in GPT-4o's error patterns implies that, although inaccurately, GPT-4o consistently reads mental-state-related information during error trials.

Together with our similarity analysis of error patterns, these findings point to a fundamental difference in how humans and GPT-4o process information in upright and inverted stimuli. For humans, inversion leads to quantitative changes in face processing[20], reducing processing efficiency[18] without altering the structure of the error space. In contrast, for GPT-4o, inversion leads to qualitative changes in the structure of its response error space. These findings speak to the decades-long debate about the face inversion effect in humans[29-31] and highlight how comparisons with LLMs may help reveal principles of information processing in both human and artificial intelligences.

The finding of a qualitative different, yet highly structured error pattern in GPT-4o for inverted stimuli has implication for human-LLM interactions. Unlike random errors, consistent errors do not cancel each other out and can introduce systematic biases. This raises the question of whether perturbations such as inversion may not only reduce GPT-4o's accuracy but also bias its reading of human mental states in systematic ways not observed in humans. For instance, during human-LLM interaction, a mental state like threatening might be consistently misinterpreted as scared, as in our data. Threatening and scared share similar emotional valence and arousal properties, yet the information they convey is different[32]. Threatening embodies a threat, while scared indicates the possibility of a threat in the environment. Systematic confusion of these mental states might lead to communicative and cooperative failures[33].

The advantage observed in our GPT-4o data for White faces in the MRMET also raises considerations about racial biases. While human participants in our sample showed no effect of stimulus race, GPT-4o performed better on images of White faces compared to Non-White faces. Prior research has identified racial biases, defined as perpetuation of racial stenotypes, in text generated by large language models[34]. Additionally, racial biases in face processing have been documented in GPT-4o and other multimodal language models[35], with bias defined as disparities in accuracy across different racial groups. An important future direction prompted by our findings will be to explore whether GPT-4o's error structure is influenced by the race of the stimulus, which could further reveal whether and how racial biases manifest in systematic errors.

The ability of GPT-4o to read complex mental states from images of the eye region has significant implications for the future of AI and human-AI interaction. As AI systems become adept at interpreting bodily cues, they could play increasingly prominent roles in fields such as mental health assessment, social robotics, and communication technologies. For instance, AI assistants equipped with advanced social cognition could provide more empathetic and personalized support to users. However, the challenges posed by a highly structured error space and the presence of racial bias underscore the need for cautious implementation.

## Methods

**Experimental models and subject details**

*Humans: RMET.* 53 human subjects were collected through the online recruitment platform Prolific to complete the RMET as a survey hosted on SoSci. The sample size was initially chosen to match the human sample size used in a previous study[6]. We recruited native English speakers between the ages of 18-70 with no history of psychiatric conditions and no history of dyslexia. Exclusion criteria were an attention-check question on the final page (which no subjects failed) and a restriction that participants could not use a smartphone or tablet to complete the task as we were concerned that the ability to easily rotate the device may impact the inversion effect (2 subjects excluded). The resulting sample was N = 51 (mean age = 33.88 years, 27 male, 24 female, 0 unknown/other; 28 White, 16 Black, 5 Asian, 2 Mixed/Other). Demographic data were not collected through self-report but were accessed through the Prolific platform. All participants provided informed consent through the online survey and received monetary compensation in return for their participation at a rate of GBP£12 per hour.

*Humans: MRMET.* Recruitment of human subjects for the MRMET used the same recruitment and exclusion criteria as the RMET. The only difference was an increase in the sample size: the calculation of information in errors involves converting the raw frequencies in a confusion matrix to joint probabilities. As a result, small sample sizes can result in bias and imprecision as a single observation can have an outsized impact on the probabilities of the matrix. To anticipate this, we conducted a sample size calculation using the MRMET validation dataset[14]. We first computed the information in errors for the full dataset, then downsampled to different sample sizes. The results of this calculation indicated that a sample size of N = 200 was sufficient to mitigate the bias and estimate the population information in errors with reasonable precision. As we had images in two orientations, we collected results from 400 participants such that each image was seen in a particular orientation by 200 subjects (mean age = 38.80 years, 179 male, 218 female, 3 unknown/other; 262 White, 69 Black, 32 Asian, 37 Mixed/Other).

*GPT-4o.* We tested OpenAI's GPT-4o, a multimodal generative pretrained transformer model with image processing capabilities that reflect the current state-of-the-art in vision benchmarks (https://openai.com/index/hello-gpt-4o/). GPT-4o debuted in May 2024 and reports being trained on material up to October 2023 (https://platform.openai.com/docs/models/gpt-4o; accessed on August 6, 2024). Data from GPT-4o (gpt-4o-2024-08-06) were collected through the OpenAI API using a custom Python script with a temperature of 0.8 and a top p of 0.9. Memory of the chat history was included within a single administration of the test (a run), but not across multiple runs, such that performance could be grouped by runs and treated as analogous to independent human subjects. The number of observations collected matched the target for human data (n = 50 for RMET; n = 400 for MRMET). Testing GPT-4o took place between 22nd August and 21st September 2024.

**Tests**

*Reading the Mind in the Eyes* (RMET). The Reading the Mind in the Eyes Test[9] is a widely used test to measure Theory of Mind and social intelligence at an advanced level in humans. The test consists of 37 black-and-white photographs (1 practice, 36 test) of the eye region of faces cropped from images appearing in magazines. Each photograph is presented with four choices of words that describe what the person in the photograph is thinking or feeling (e.g. concerned, panicked, friendly, serious), of which one word is the target (correct) answer. Participants are instructed to choose the one word that best describes what the person in the photograph is thinking or feeling, and to answer as quickly and accurately as possible.

*Multiracial Reading the Mind in the Eyes* (MRMET). The Multiracial Reading the Mind in the Eyes Test[14] is a recent, updated version of the classic RMET that incorporates racially inclusive stimuli, nongendered answer choices, ground-truth referenced answers, and a more accessible vocabulary. The test also includes 38 (1 practice, 37 test) high resolution, full colour photographs of unfamiliar identities. This test has been fully validated and found to be fully interchangeable with the RMET across major psychometric indices.

**Testing protocol**
For each test, images were presented sequentially with the question, "What emotion do the eyes in the image show? Choose only one emotion from the following options:" followed by the four options for that item listed from (a) to (d). The order of images was fixed in line with the published standards of the two tests, as was the order of alternatives for each item. For humans, orientation of the images was manipulated within-subjects such that participants saw upright or inverted images in a pseudorandomised order. The first (practice) image was always presented upright, but the presentation of later images was counterbalanced across participants. The options were presented underneath the image as radio buttons, meaning that participants could only select one of the four options. For GPT-4o, orientation was manipulated between-runs such that all images within a given test run would be presented upright or inverted. Responses from GPT-4o were open-ended text responses, which were subsequently coded by experimenters to record which of the four items has been identified. Uncodable responses (such as "I cannot tell what the emotion is from the eyes alone") were marked as NA. A small number of runs that consisted only of uncodable responses were discarded and recollected. Part of the RMET includes a glossary of definitions that are at the participant's disposal in case they do not recognise or understand a given word in the test. This glossary was not considered necessary for testing GPT-4o, but it was coded into the survey for collecting human data such that participants could see the definition of a choice by hovering their mouse over the word. The MRMET was designed to use a more accessible vocabulary than the RMET and so does not require a glossary of terms.

**Quantification and statistical analysis**
**Logistic Mixed Effects Models for assessing statistical differences in accuracy**. Responses to all items were coded with a numeric value from 1 to 4 to indicate the reported label, and accuracy was determined by the match between this value and the designated correct label (1 correct; 0 incorrect). Accuracy for each response was then analysed using a Logistic Mixed Effects Models[36,37]. We considered single-trial correct response (0,1) as the dependent variable, model (human, GPT-4o) and orientation (upright, inverted) as fixed effects, and observation ID (subject for humans, run for GPT-4o) as a random intercept. The significance of all fixed effects was assessed by conducting likelihood-ratio tests (LRT) between mixed models differing only in the presence or absence of the given predictor[38]. Interactions were examined using the R package *emmeans*[39], which provides estimated marginal means for predicted probabilities in logistic models. Selection of the random-effects structure was based on the Bayes information criterion (BIC)[40]. The model including the random effects structure was chosen over a minimal model without random structure when BIC was lower. Models were built using the *glmer* function from the R package *lme4*[41,42].

**Error concentration**. Responses to all items on the MRMET were recoded to assign a numeric value to each label in the dataset. In all, there were 40 unique mental state labels included in the test. Presented categories (i.e., the labels that served as correct answers to the questions) were assigned codes 1-17, and alternative options (the labels that never appeared as the correct answer) were assigned codes 18-40. Some presented category labels repeated within the test but were always paired with the same three alternatives. Two data vectors were computed using these labels: one designating the presented (correct) response for each question, and one designating the reported response for each question. These vectors were used to construct a confusion matrix[23,24], $P(s'|s)$, reporting the empirical conditional probability that a mental state $s'$ is reported given that mental state $s$ was presented in the same trial. This was computed by counting the occurrences of each reported mental states $s'$ and dividing them by the total number of presentations of mental state $s$. The sum of the on-diagonal cells, where the reported stimulus matched the presented stimulus, weighted by the probability of presentation of each mental state $s$, equals the reporting accuracy (fraction correct) $f_{cor}$.

To quantify how errors were distributed across the three incorrect alternatives, we computed the conditional probability of reporting each alternative (corresponding to the elements of the confusion matrix for the empirically possible alternatives for each presented mental state) for each of the 17 presented mental states. We ranked these probabilities in descending order, from most to less probable and then we averaged them across all presented mental states for which at least one error occurred. We indicate the so obtained probability by $P(x)$, where $x = 1:3$ indexes the three alternatives ordered in terms of descending probabilities. The concentration of errors across alternative probabilities was quantified by fitting, separately for each condition, $P(x)$ with an exponential decay function:

$$P(x) = \frac{\exp(-\lambda x)}{(\sum_{x=1}^{3} \exp(-\lambda x))} \quad (1)$$

where $\lambda$ describes the concentration index (the larger lamba, the more concentrated on the most likely alternative are the errors). We selected the exponential decay model because it fit the data reasonably well (see Figure 2B) and offers a straightforward way to quantify error concentration using a single parameter. Exponential decay models were fitted using MATLAB's fit function with nonlinear least squares optimization and a randomly initialized starting point for parameter estimation.

**Structure similarity.** To quantify the similarity in patterns of correct and incorrect reports between models and orientation, we flattened each conditional confusion matrix into a vector and computed the Spearman correlation coefficient between human and GPT-4o upright and inverted vectors separately for correct classifications (on-diagonal elements in the confusion matrices) and misclassifications (off-diagonal elements). To prevent correlations from being artificially inflated by zero entries — where stimuli or alternative options were not presented — we restricted the analysis to data corresponding to experimentally presented mental states. Structure similarity was defined as the Spearman correlation coefficient between the two vectors. We chose Spearman correlation due to its generality and robustness for computation of similarities and dissimilarities.

**Information in errors**. For each confusion matrix, we computed the mutual information in the confusion matrix using Shannon's formula[23,24]:

$$I = \sum_{s,s'\in S} P(s)P(s'|s)\log_2 \frac{P(s'|s)}{P(s')} \qquad (2)$$

where in the above $P(s'|s)$ is the confusion matrix quantifying the empirical conditional probability that a mental state $s'$ was reported given that mental state $s$ was presented in the same trial, and $P(s), P(s')$ are the marginal probabilities of presented and of reported mental states. The mutual information in the confusion matrix (I) captures the total information carried by both correct responses and errors. To estimate the information contributed specifically by the error structure (beyond that contributed by the fraction correct), we first computed $I_{flat}$, the information contained in a confusion matrix where correct responses (the diagonal elements) are preserved, but the information added by the error structure is disrupted by flattening the error distribution (off-diagonal elements) for each presented mental state. The difference between I and $I_{flat}$ quantifies the information specifically contributed by the error structure given the fraction correct.

To normalize this information (expressed by I - $I_{flat}$), we next computed $I_{conc}$, the information contributed when the confusion matrix exhibits maximal structure — that is, when errors are maximally concentrated on the response option, from the empirically possible alternatives, that maximally differentiates between the presented mental states (see below). The difference between $I_{conc}$ and $I_{flat}$ quantifies the maximal information that could be added by the error structure given the fraction correct.

The normalized information metric, $I_{errors}$, is obtained by dividing the information specifically contributed by the error structure, I - $I_{flat}$, by the maximal information that error structure could contribute, $I_{conc}$ - $I_{flat}$,[26,43,44]:

$$I_{errors} = \frac{I - I_{flat}}{I_{conc} - I_{flat}} \qquad (3)$$

A value of 0 corresponds to information equalling that of the flat error distribution, while a value of 1 corresponds to information equalling that of the fully concentrated, maximally informative, error distribution.

The amount of information in a maximally concentrated error distribution can vary depending on which alternative the errors are concentrated on for each presented mental state[26]. This is because different alternatives may differentiate between presented mental states to different degrees. To numerically determine $I_{conc}$, we randomly permuted 1 million times for each presented mental state the identity of the experimentally possible alternative containing all the errors. We computed $I_{conc}$ as the maximum of confusion matrix information over all permutations — which correspond to concentrating errors on the alternative, among the experimentally possible ones, that maximally differentiates between the presented mental states. We verified that calculating $I_{conc}$ as the average over all permutations, or by concentrating all errors on the experimentally determined most likely alternative without permutation, yielded the same pattern of results across models and orientations.

The mean and SE of the concentration index and information in errors were computed by bootstrapping[45] and were then used to compute comparison significance statistics using unpaired t-tests with two-tailed Holm-Bonferroni corrected *P* values.

**Matching performance**. To control for the influence of variability of performance on the differences in concentration index and information in errors, we conducted a control analysis using the MRMET validation dataset[14]. This dataset consists of responses from 9,925 human subjects who completed the MRMET (upright only). We sampled 200 subjects from this dataset to approximate the mean and standard deviation of fraction correct of the GPT-4o sample using a hill-climbing optimisation algorithm[46], a stochastic global search optimisation strategy that initialises a randomly drawn subset of subjects and then iteratively swaps one out and takes the result with the smaller distance from the target distribution for 10,000 iterations. Due to the large sample, there were multiple combinations of subjects who could approximate this target distribution, and initial computations revealed variability in the computed information, so we ran this optimisation 500 times to create 500 subsets of N = 200 human subjects whose performance closely matched that of GPT-4o. Information was calculated for each of these subsets, and the distribution was inspected to see if it was centred closer to the observed N = 200 human dataset that we collected or closer to the GPT-4o dataset whose performance it matched.

**Conventions for *P* values and confidence intervals.** The *P* values of all reported statistical comparisons are two-tailed and Holm-Bonferroni correction for multiple comparisons unless otherwise stated. 95% confidence intervals reported in Figures 1-3 correspond to multiplying the SE by 1.96 and adding/subtracting this margin from the means.

**Data and code availability**
The data and code supporting this study will be made publicly accessible upon journal publication.

**Acknowledgements**
This work is supported by the European Commission through Project ASTOUND (101071191 — HORIZON-EIC-2021-PATHFINDERCHALLENGES-01, A.R., F.M., G.M., S.P., C.B.). CY was supported by a China Scholarship Council Fellowship. The funders had no role in study design, data collection and analysis, decision to publish or preparation of the manuscript.

**Author contributions**
J.W.A.S., O.P., E.S., M.S.A.G., and C.B. conceived the study. J.W.A.S., O.P., E.S., G.M., and K.S. performed the experiments including preparation of the scripts, data collection with humans and GPT models, response coding, and curation of the dataset. All authors contributed materials and methods. S.P. designed the analyses with contributions from J.W.A.S. and C.B. J.W.A.S. and S.P. performed the analyses. C.B. wrote the manuscript with contributions from J.W.A.S. and S.P. All authors contributed to the interpretation and editing of the manuscript. C.B. supervised the work. A.R., F.M., G.M., S.P. and C.B. acquired the funding. O.P. and E.S. contributed equally to the work.